\documentclass[aps,prb,twocolumn,superscriptaddress,floatfix,letter]{revtex4-1}
\usepackage{epsfig,amsmath,amssymb,color,dsfont,physics}
\definecolor{jadclr}{rgb}{0,0.5,0}
\definecolor{jadcolor}{rgb}{0.5,0,0}
\usepackage[bookmarks=true,colorlinks,linkcolor=jadclr,urlcolor=jadcolor,citecolor=jadcolor]{hyperref}
\usepackage{cleveref}
\usepackage{bm}
\usepackage{booktabs}
\usepackage{mathrsfs}
\usepackage{array}

\def\8{\infty}
\def\oh{\frac{1}{2}}

\def\undertext#1{\vtop{\hbox{#1}\kern 1pt \hrule}}

\def\tr{\hbox{tr}\,}

\def\be{\begin{equation}}
\def\ee{\end{equation}}
\def\bea{\begin{eqnarray} & &}
\def\eea{\end{eqnarray}}

\def\rf#1{(\ref{#1})}

\def\rfs#1{Eq.~\rf{#1}}

\makeatletter
\newcommand{\thickhline}{%
	\noalign {\ifnum 0=`}\fi \hrule height 1pt
	\futurelet \reserved@a \@xhline
}
\newcolumntype{"}{@{\hskip\tabcolsep\vrule width 1pt\hskip\tabcolsep}}
\makeatother
\begin{document}


\title{Dynamical quantum phase transitions in many-body localized systems}

\author{Jad C.~Halimeh}
\affiliation{Max Planck Institute for the Physics of Complex Systems, N\"othnitzer Stra\ss e 38, 01187 Dresden, Germany}
\affiliation{Physics Department, Technical University of Munich, 85747 Garching, Germany}
\author{Nikolay Yegovtsev}
\affiliation{Department of Physics and Center for Theory of Quantum Matter, University of Colorado, Boulder, Colorado 80309, USA}
\author{Victor Gurarie}
\affiliation{Department of Physics and Center for Theory of Quantum Matter, University of Colorado, Boulder, Colorado 80309, USA}

\date{\today}

\begin{abstract}
We investigate dynamical quantum phase transitions in disordered quantum many-body models that can support many-body localized phases. Employing $l$-bits formalism, we lay out the conditions for which singularities indicative of the transitions appear in the context of many-body localization. Using the combination of the mapping onto $l$-bits and exact diagonalization results, we explicitly demonstrate the presence of these singularities for a candidate model that features many-body localization. Our work paves the way for understanding dynamical quantum phase transitions in the context of many-body localization, and elucidating whether different phases of the latter can be detected from analyzing the former. The results presented are experimentally accessible with state-of-the-art ultracold-atom and ion-trap setups.
\end{abstract}

\maketitle

\section{Introduction}
After the establishment of universality and scaling in equilibrium, \cite{Cardy_book,Sachdev_book} particularly with the invention of the renormalization group,\cite{Wilson1971a,Wilson1971b} their extension to out-of-equilibrium systems has become a major area of research in physics. 
Notwithstanding considerable progress for classical out-of-equilibrium systems,\cite{Hohenberg1977} out-of-equilibrium dynamics for quantum
systems remains an active frontier. In recent years, the field of dynamical quantum phase transitions\cite{Heyl2013,Heyl2014,Heyl2015} (DQPT) has 
shed new light on the out-of-equilibrium quantum many-body criticality. The concept of DQPT relies on an intuitive analogy between the thermal partition function and the overlap of an initial state and its time-evolved self in the wake of a quench, referred to as the \textit{Loschmidt amplitude} in this context. One can thus construct the Loschmidt return rate, which is proportional to the logarithm of this overlap, as a dynamical analog of the thermal free energy, with evolution time standing for complexified inverse temperature. As such, nonanalyticities in the return rate would occur at \textit{critical times} much the same way a thermal phase transition manifests itself as a nonanalyticity in the thermal free energy at a critical temperature. 


Even though quite a few of the original questions raised at the onset of the field of DQPT have now been answered, in the quest to attain these answers far more questions have arisen.\cite{Heyl_review} For example, in the initial study of quenches in the one-dimensional transverse-field Ising model,\cite{Heyl2013} the singularities in the return rate were found only after quenches across the equilibrium quantum critical point.\cite{Heyl_review} Soon afterwards it became clear that this was not the necessary condition for DQPT. Rather, one can define
the concept of a \textit{dynamical} quantum critical point, which may in some cases coincide with the equilibrium one, that separates quenches with DQPT from those without.\cite{Karrasch2013,Andraschko2014,Vajna2014} 

Yet later it was observed that in long-range quantum spin chains with power-law interactions,\cite{Halimeh2017a,Zauner2017} the DQPT occur regardless of the quench. Nevertheless, a meaningful dynamical critical point can still be defined, this time separating qualitatively
different types of evolution after the quench, which notably depends on whether the order parameter oscillates about zero in the time-evolved quenched system.\cite{Halimeh2017a,Zauner2017,Homrighausen2017,Lang2018a,Lang2018b} This in turn spurred further study of the
steady states reached after the quench, their relationship with the DQPT occurring in the evolution towards them, and the effects of
quasiparticle excitations, which is still ongoing.\cite{Halimeh2018,Hashizume2018,Vanderstraeten2018,Liu2018,Jafari2019,Defenu2019}

The large body of theoretical work on DQPT in closed clean quantum many-body systems in the wake of a quantum quench has recently been supported by experimental realizations in ion-trap\cite{Jurcevic:2017be} and ultracold-atom\cite{Flaeschner2018} setups. Naturally, the robustness of DQPT has since been investigated in other paradigms of quantum many-body physics, such as Floquet systems\cite{Kosior2018a,Kosior2018b,Yang2019} where a novel type of ``Floquet singularities'' appear as intrinsic features of periodic time modulation, and also in disordered systems,\cite{Gurarie2018} where singularities arise in the return rate of random-field classical Ising models. This latter result directly inspires the investigation of DQPT in many-body localized (MBL) quantum systems\cite{Huse2015,Altman2015,Abanin2017} as the next frontier in this field. Indeed, MBL systems are drastically alien to the models that have thus far been utilized in the study of DQPT. On the one hand, so-called \textit{emergent integrability} due to the failure of MBL systems to thermalize after a quantum quench may perhaps allow one to consider that DQPT behavior should not be ruled out given its prevalence in clean integrable models. On the other hand, how the emergent exotic MBL phases, if at all, are related to possible DQPT nonanalyticities is a completely open question. Furthermore, since the exact origin of DQPT is still not fully understood, with a quasiparticle origin thus far the most convincing argument,\cite{Halimeh2018,Jafari2019,Defenu2019} studying DQPT in MBL systems can possibly shed further light on what exactly the necessary and sufficient conditions are for singularities to arise in the return rate.

The Loschmidt amplitude is intimately connected to the partition function of the corresponding system
 evaluated at imaginary temperature as we elaborate below.
In turn, the imaginary-temperature partition function measures the spectral form factor, or the correlations between energy levels in a quantum system.\cite{Gurarie2018} DQPT would then be equivalent to having strong correlations between energy levels across significant energy intervals much larger than the level spacing. We would expect those correlations in integrable and close-to-integrable systems, as well as in some MBL systems, but not in chaotic quantum systems.  This line of inquiry further motivates studying MBL systems for signatures of DQPT in them.
 
In this work, we provide analytic and numerical evidence showing that quantum many-body systems that support MBL can indeed show rich DQPT behavior. 

The rest of the paper is organized as follows: In Sec.~\ref{sec:PartFunc}, we review the imaginary-temperature boundary and full partition functions, and present a candidate quantum MBL model in whose partition functions singularities arise. In Sec.~\ref{sec:lbits}, we use the $l$-bits formalism to determine the conditions under which disordered quantum models will exhibit singularities in their imaginary-temperature full partition function. We present and discuss the DQPT in our candidate model in Sec.~\ref{sec:results}. We conclude in Sec.~\ref{sec:conc} and provide an outlook on future work. Additional details on the $l$-bits analysis are provided in Appendix~\ref{sec:app}.

\section{Partition functions and models}\label{sec:PartFunc}
We would like to study the probability that a quantum system placed initially in a state $\ket{\psi}$, after evolving for time $t$, returns back to its original state $\ket{\psi}$. This quantity can 
be written as
\begin{align} 
	Z(t) = \bra{\psi} e^{-i H t} \ket{\psi} = \sum_{\alpha} \left| c_\alpha \right|^2 e^{-i E_\alpha t}.
\end{align}
Here $c_\alpha$ are the overlap coefficients between the state $\ket{\psi}$ and the eigenstates $\ket{\psi_\alpha}$ of the Hamiltonian $H$ corresponding
to the energy levels $E_\alpha$. $Z(t)$ is a boundary partition function construed in the seminal DQPT work of Ref.~\onlinecite{Heyl2013} as a dynamical analog of the thermal partition function. Consequently, the Loschmidt return rate,
\be \label{eq:lerr} \lambda(t)=-\frac{1}{N}\ln|Z(t)|^2, \ee 
is a dynamical analog of the thermal free energy, with $N$ standing for system size. 

General principles of statistical physics dictate that for large systems $Z(t)$ may be well approximated by the thermal finite-time partition function 
\be {\cal Z} (t,T)= \sum_\alpha e^{-E_\alpha \left(it + \frac 1 T \right)},
\ee with the appropriately chosen temperature $T$, such that
\be \left< \psi \right| H \left| \psi \right> =\frac{\sum_\alpha E_\alpha e^{- \frac{E_\alpha}{T}}}{\sum_\alpha e^{- \frac{E_\alpha}{T}}}.
\ee 
Among these, the $T\to\infty$ partition function is especially suitable for study: 
\begin{align} 
	\mathcal{Z}(t) =\lim_{T \rightarrow \infty} {\cal Z}(t,T) =  \sum_\alpha e^{-i E_\alpha t}.
\end{align}
Depending on the choice of $\ket{\psi}$, in some cases $Z(t)$ and ${\cal Z}(t)$ may even coincide, specifically when $\left| \psi \right>$ 
is proportional to the sum 
$\sum_\alpha \left| \psi_\alpha \right>$. 

At the same time, $\left| {\cal Z}\right|^2$ is the so-called spectral form factor of the system,
\begin{align} \left| \mathcal{Z}(t) \right|^2 = \sum_{\alpha,\beta} e^{i (E_\beta-E_\alpha) t}.
\end{align} 
The Fourier transform of the spectral form factor  over time $t$ is simply the correlation between the density of states at energies separated by
a certain energy interval,
\begin{align} && \int \frac{dt}{2\pi} \left| {\cal Z}(t) \right|^2 e^{i \omega t} =\sum_{\alpha \beta} \delta(\omega-E_\beta+E_\alpha)
\cr
&&= \int dE \, \rho\left( E + \omega/2 \right) \, \rho\left(E- \omega / 2 \right). \end{align}
Here,
\begin{align} \rho(E) = \sum_\alpha \delta \left( E - E_\alpha \right)
\end{align} is the density of states.
A nonanalyticity in ${\cal Z}(t)$ at a certain time $t_c$ is therefore a reflection of strong correlations existing between energy levels at the energy interval
$\omega_c \sim 1/t_c$. 


It might be quite surprising to expect these correlations to persist over energy intervals far larger than the level spacing for an interacting quantum many-body system. Indeed, an extreme example of a generic quantum system is provided by the Random Matrix Theory (RMT). Within that theory, average spectral form factors are known to have a single nonanaliticity at $t=t_c$ equal to the inverse average level spacing.\cite{Mehta} Level spacings in generic quantum many-body theories are exponentially small in system size, resulting in an exponentially large time at which the potential nonanalytic behavior occurs. These types of singularities would then be nonobservable.\cite{footnote}


In view of this argument, it is quite remarkable that a number of models already discussed in the literature display singularities at times of the order of inverse couplings in the Hamiltonian. In other words, in these models the density of energy levels correlates across energy intervals that cover a large number of energy levels. Some of these models are integrable. It is fair to declare that in this context the existence of subtle correlations in energy levels is not surprising. Some others are nonintegrable, however. 

We will not attempt here to analyze the origin of energy level correlations in these models.  Instead we would like to look at systems which look as different from RMT as possible. RMT is supposed to represent generic ``chaotic" quantum systems with Wigner-Dyson level statistics. Quite distinct from these are systems with Poisson level statistics, including integrable models. A whole separate class of quantum systems with Poisson level statistics are MBL models. Here, we would like to discuss the appearance of singularities
in ${\cal Z}(t)$ and $Z(t)$ of MBL systems. 


%



Let us concentrate on models of one-dimensional spin-$1/2$ chains of length $N$ with various random interaction terms in the Hamiltonian. Those are known to be many-body localized when disorder is strong enough.\cite{Alet2018} An example of such an MBL model can be the Ising model with random bonds and random fields
\begin{align} \label{eq:mb1} H = \sum_{n=1}^{N-1} J_n \sigma^z_n \sigma^z_{n+1} + \sum_{n=1}^N \left( h^\perp_n \sigma^x_n + h^\parallel_n \sigma^z_n \right).
\end{align}
Here $\sigma^z_n$ and $\sigma^x_n$ are Pauli matrix operators acting on spins which reside on sites $n$ of a one-dimensional lattice. All $J_n$, $h^\perp_n$ and $h^\parallel_n$ are independent random variables. 

We shall present a series of analytic and numerical arguments supporting the idea that a certain class of these one-dimensional spin-$1/2$ MBL models will have singularities in their ${\cal Z}(t)$. In particular, we will demonstrate that while ${\cal Z}(t)$ is not singular in the model given by~\eqref{eq:mb1}, the following
MBL Hamiltonian 
\begin{align} \label{eq:mb2} H = h \sum_{n=1}^N  \sigma^z_n + \sum_{n=1}^{N-2} J_n \sigma^x_n \sigma^x_{n+1} \sigma^x_{n+2},
\end{align}
with $J_n$ being independent random variables, features singularities in its ${\cal Z}(t)$ as well as in its return probability $Z(t)$ defined for suitable initial states $\left| \psi \right>$.

\section{$l$-bits and return rate}\label{sec:lbits}
Locally conserved quantities called $l$-bits in the context of many-body localization play an important role in our arguments, so let us
review their definition.  As nicely argued in Ref.~\onlinecite{Abanin2015} for any spin-$1/2$ Hamiltonian $H$ it is always possible to construct a set of mutually commuting operators $\tau_n^z$ whose square is identity $(\tau_n^z)^2=1$, which commute with any spin-$1/2$ Hamiltonian $H$. A straightforward technique to do that would be to diagonalize the Hamiltonian $H$ such as the one above. In other words we use the natural basis of spin tensor product states and write  $H = U \Lambda U^\dagger$ where $U$ is a unitary matrix mapping the product state basis onto the basis of eigenstates of $H$, and $\Lambda$ is diagonal. Define then \begin{align} \tau_n^z =  U\sigma^z_n U^\dagger.
\end{align} By construction, these $\tau_n^z$ all square to 1, commute with each other, and all commute with the Hamiltonian. 
Furthermore, it is always possible to rewrite the Hamiltonian in terms of $\tau_n^z$ according to
 \begin{eqnarray} \label{eq:lbits} H &=& K^{(0)}+\sum_n K^{(1)}_n \tau_n^z + \sum_{nm} K^{(2)}_{nm} \tau_n^z \tau^z_{m}  
 \cr && +\sum_{nml} K^{(3)}_{nml} \tau_n^z \tau^z_{m} \tau^z_l + \dots.
\end{eqnarray}
It is obvious that the total number of the coefficients $K$ in~\eqref{eq:lbits} is $2^N$, same as the number of eigenvalues of $H$, so they are sufficient to fully parametrize the Hamiltonian. These coefficients can all be found using
\begin{align} \label{eq:lcoef} K^{s}_{n_1 \dots n_s} = 2^{-N} \tr \left( H \tau^z_{n_1} \dots \tau^z_{n_s}  \right).
\end{align}

By itself this construction is too general to be of much use. However, for MBL Hamiltonians the expansion~\eqref{eq:lbits} simplifies 
significantly. $U$ is defined up to the permutations of eigenvalues of the Hamiltonian. It can be argued that with the appropriate choice of $U$, the operators $\tau_n^z$ become local, that is, they can be written as a linear combination of terms involving products of spin operators $\sigma^x_m$, $\sigma^y_m$, and $\sigma^z_m$ on sites $m$ nearby $n$. At the same time, the series~\eqref{eq:lbits} also become local, in that the magnitudes of the coefficients $K^{(s)}_{n_1 \dots n_s}$ drop off exponentially with $s$, as well as with the increasing separations between the lattice sites $n_1$, $n_2$, $\dots$, $n_s$ in these coefficients. In this regime, $\tau_n^z$ are usually referred to as $l$-bits, with $l$ standing for localized. The coefficients $K^{(s)}$ are random, being some complicated combination of the random interaction coefficients of the original Hamiltonian. 

The description in terms of $l$-bits allows us to rewrite the imaginary-temperature partition function in a very straightforward fashion:
\begin{eqnarray} \label{eq:classic} {\cal Z}(t) & = & e^{-i t K^{(0)}}  \times \cr && \sum_{\tau = \pm 1} e^{-i t \sum_n K^{(1)}_n \tau_n - i t \sum_{nm} K^{(2)}_{nm} \tau_n \tau_m - \dots} .
\end{eqnarray}
Here, the variables $\tau_n$ are eigenvalues of the $l$-bits $\tau^z_n$ taking values $\pm 1$. We can use~\eqref{eq:classic} as a starting point to calculate
partition functions of one-dimensional spin-$1/2$ MBL models. Furthermore, since in the MBL phase the coefficients $K^{(s)}$ quickly go to zero with increasing $s$, it is sufficient to retain just a few terms in the series in the exponential of~\eqref{eq:classic}, which is what we will rely on below. 

Since the coefficients $K^{(s)}_{n_1 \dots n_s}$ are random, it is important to decide which quantities can be averaged over many sets of these random coefficients. The quantity ${\cal Z}(t)$ is not self-averaging. This means that computing it over a particular set of $K^{(s)}_{n_1 \dots n_s}$ taken from the original randomly generated interaction coefficients in the MBL Hamiltonian is not the same as averaging it over many realizations of them. On the contrary, the return rate \begin{align}\label{eq:PFRR} r(t) =-\frac{1}{N}\ln\frac{\left| 
{\cal Z} \right|^2}{2^{2N}}, \end{align} defined analogously to the Loschmidt return rate~\eqref{eq:lerr}, is a self-averaging quantity well defined in the ``thermodynamic limit" of infinite chain $N \rightarrow \infty$. It is $r(t)$ that can be averaged over many realizations. Its average value should coincide with its typical value computed for a  particular set of $K^{(s)}_{n_1 \dots n_s}$. Various constants present in the definition of $r(t)$ are there  merely for normalization
purposes.

In the work of Ref.~\onlinecite{Gurarie2018}, the partition function was evaluated for the model given by~\eqref{eq:classic} with $K^{(1)}_n = h_n$ random and 
$K^{(2)}_{nm} =( \delta_{n, m-1} + \delta_{n,m+1} ) J$ representing constant nearest-neighbor interactions. In other words, the following model
was studied:
\begin{align} \label{eq:old} {\cal Z} (t) = \sum_{\tau = \pm 1} e^{- i t \sum_n h_n \tau_n - i t J \sum_{n=1}^{N-1} \tau_n \tau_{n+1}},
\end{align} with $h_n$ random independent variables.  It was found, both numerically and analytically, that the resulting function $r(t)$ is  singular at the points in time $t_n = n \pi/(2J)$, with $n$ an arbitrary integer. Each singularity was found to be of the type $|t-t_n| \ln |t-t_n|$. 

At the same time, numerical evidence indicated that if $J$ in~\eqref{eq:old} is promoted to a bond-dependent random variable $J_n$, then $r(t)$ is featureless and has no singularities. We expect it to be true generally: a generic model~\eqref{eq:classic} with all the coefficients being 
independently random  variables will feature no singularities and in fact for large system size its ${\cal Z}(t)$ should self-average to a time-independent constant.

We would like to generalize beyond~\eqref{eq:old}. In the next section we present evidence supporting the conjecture that  as long as there is at least one nonrandom interaction coefficient in~\eqref{eq:classic}, with the
rest being random, the return rate $r(t)$ always features singularities, at positions $t_n=n \pi/(2K)$, where $n$ is an arbitrary integer and $K$ is the nonrandom interaction coefficient in~\eqref{eq:classic}. 

A generic MBL model can be expected to map onto~\eqref{eq:classic} with all interaction coefficients random. However, it is also natural to expect
that models should exist whose mapping to~\eqref{eq:classic} feature at least one nonrandom coefficient. Those models would then have singularities in their return rate. Furthermore, a possibility  should not be entirely discounted that~\eqref{eq:classic} with all coefficients random but correlated in a certain
way would also feature singularities, and that MBL models exist which map onto these kinds of $l$-bit models. In the next section we present further examples of random $l$-bit models with singularities in their return rate, and demonstrate the existence
of singularities in the MBL model given by~\eqref{eq:mb2}.

\section{Results and discussion}\label{sec:results}
\begin{figure}[tb]
	\centering
	\includegraphics[width=0.46\textwidth]{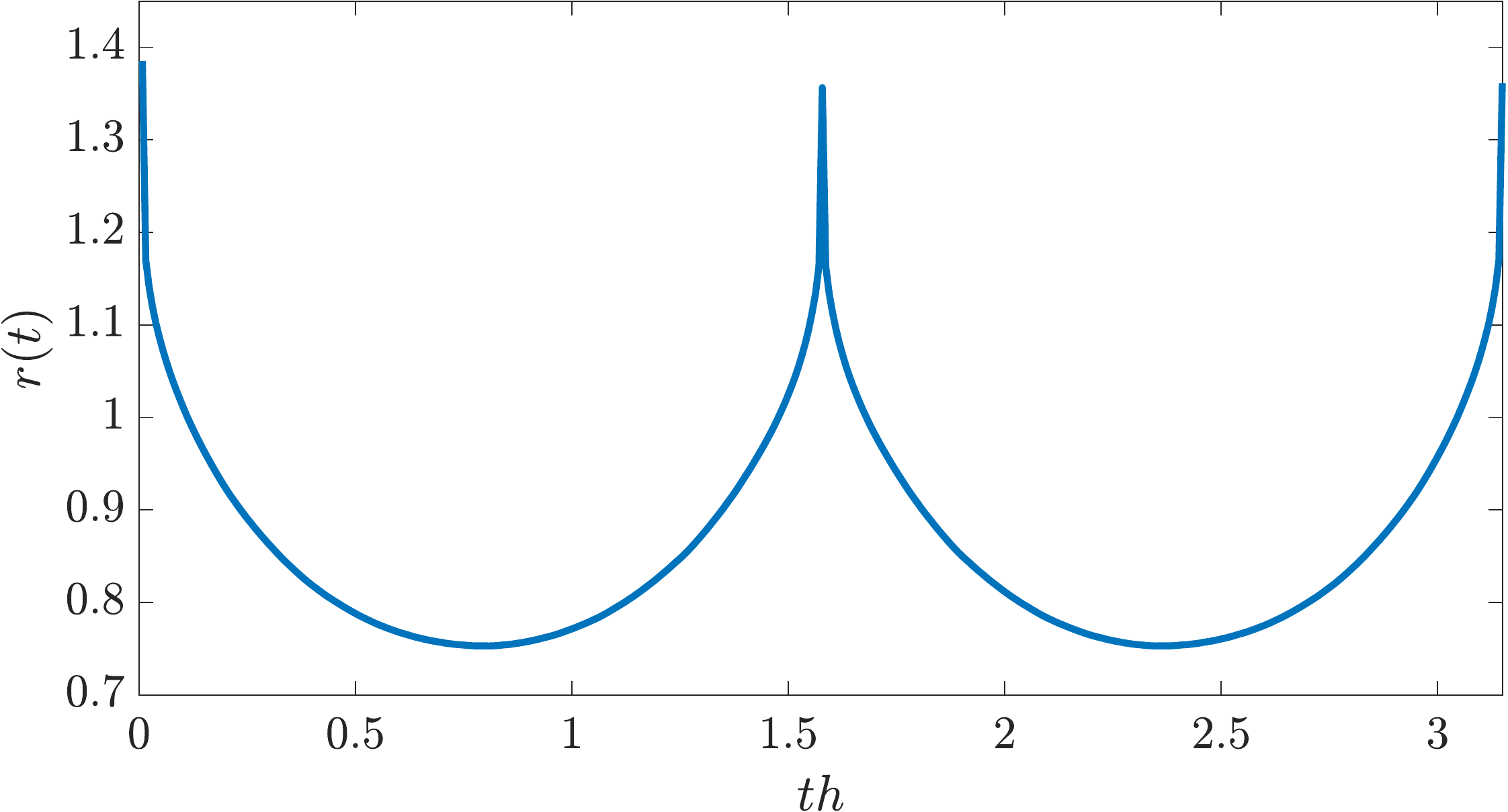}
	\caption{Return rate $r(t)$ evaluated for~\eqref{eq:randombond2} for $N=45000$ showing clear singularities. Beyond the largest time shown here, the return rate repeats periodically.}
	\label{introfig} 
\end{figure}
\begin{figure}[tb]
	\centering
	\includegraphics[width=0.46\textwidth]{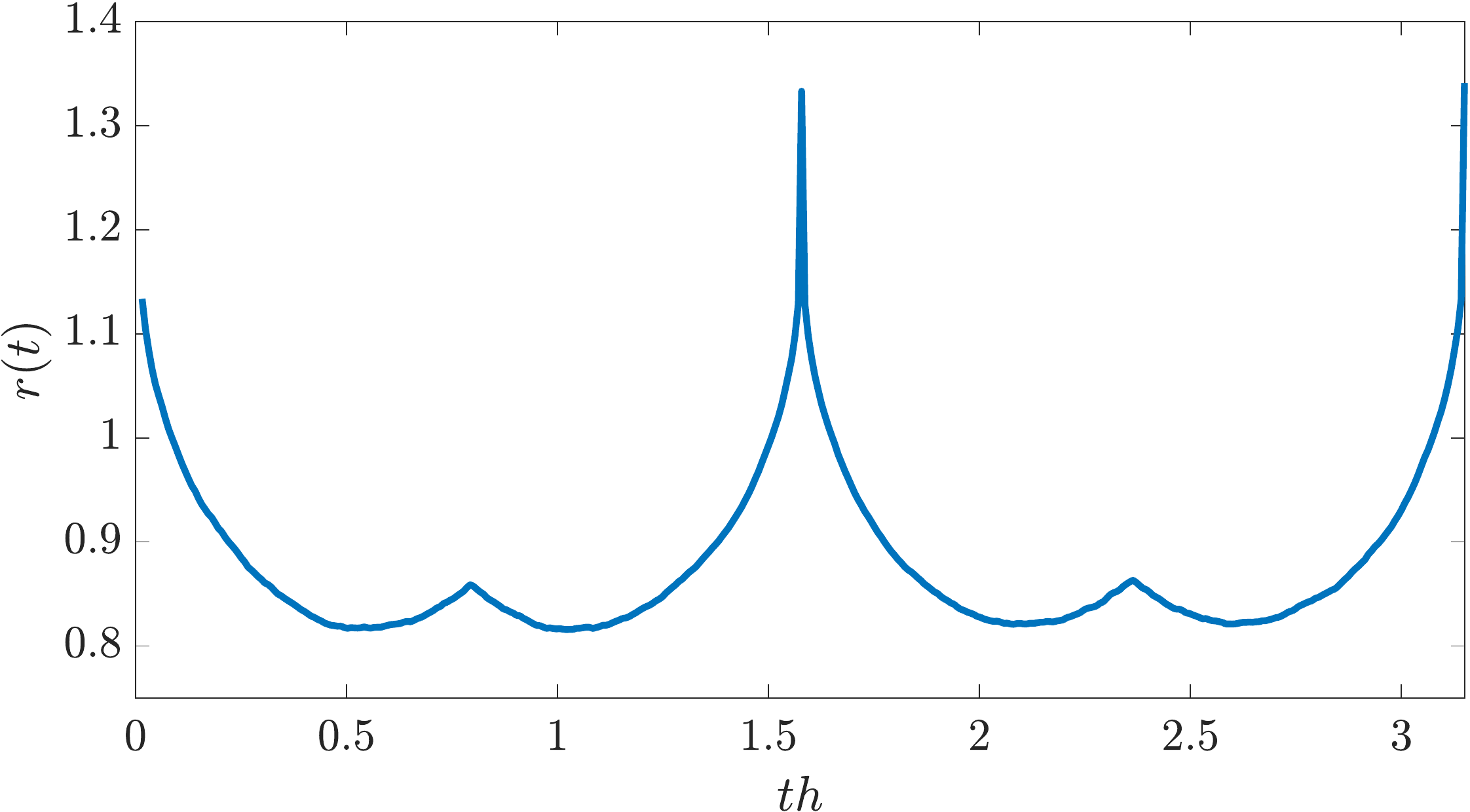}
	\caption{Same as Fig.~\ref{introfig} but for~\eqref{eq:randombond3}, again with $N=45000$. $r(t)$ appears qualitatively different from
		that in Fig.~\ref{introfig} but features the same singularities at $th=\pi n/2$. The return rate repeats periodically beyond the largest time shown here.}
	\label{introfig2} 
\end{figure}
An important remark which simplifies further analysis lies in the observation, already discussed in Ref.~\onlinecite{Gurarie2018}, that a random variable $e^{-i t h_n  \sigma}$, for $\sigma = \pm 1$ and $t$ sufficiently large, regardless of the probability distribution for $h_n$, is well approximated by the variable $e^{-i f_n \sigma}$ where $f_n$ is now taken as uniformly distributed on the interval $f_n \in [-\pi, \pi]$. This should be fairly obvious: the latter variable is uniformly distributed over a unit circle in the complex plane, while the former approaches this distribution at large enough $t$. 

We would now like to illustrate the principle of a single nonrandom coefficient in~\eqref{eq:classic} leading to singularities by considering for example
\begin{align}  \label{eq:randombond1} {\cal Z}(t) = \sum_{\tau = \pm 1} e^{- i t h \sum_n  \tau_n - i t  \sum_{n=1}^{N-1} J_n \tau_n \tau_{n+1}},
\end{align}
with $J_n$ random. In line with the observation in the previous paragraph, we instead study
\begin{align} \label{eq:randombond2}   {\cal Z}(t) = \sum_{\tau= \pm 1} e^{- i t h \sum_n  \tau_n - i   \sum_{n=1}^{N-1} f_n \tau_n \tau_{n+1}},
\end{align}
with $f_n$ independent random variables uniformly distributed over the interval $[-\pi, \pi]$. 
The advantage of this representation of ${\cal Z}(t)$ is that~\eqref{eq:randombond2} is automatically periodic in $t$, as is clear by inspection. At the same time, the original model~\eqref{eq:randombond1} approaches~\eqref{eq:randombond2} at sufficiently large $t$, $t \gtrsim 2\pi/h$. This can also be easily verified numerically. 

Fig.~\ref{introfig} shows $r(t)$ evaluated for~\eqref{eq:randombond2} for $N=45000$ sites and $th$ ranging from $0$ to $\pi$. $r(t)$ continues periodically beyond the displayed range of $t$. To produce this result, we took advantage of the standard transfer matrix calculation 
of the partition function, which allowed us to easily go to fairly large system sizes. The self-averaging property of $r(t)$ is obvious in Fig.~\ref{introfig}: even though only one realization of disorder is taken, the curve is smooth.

Similarly we can evaluate, for example, the partition function
\begin{align} \label{eq:randombond3}   {\cal Z}(t) = \sum_{\sigma = \pm 1} e^{- i t h \sum_n  \tau_n - i   \sum_{n=1}^{N-2} f_n \tau_n \tau_{n+1}
\tau_{n+2}},
\end{align} with the result shown in Fig.~\ref{introfig2}. 

This shows a new feature at $t h = \pi/4 + \pi n/2$, however the more interesting singularity still remains at $t h =  \pi n/2$. All this is compatible with the observation that one nonrandom coefficient in~\eqref{eq:classic} is sufficient to generate periodically repeating singularities in ${\cal Z}(t)$. 

\begin{figure}[tb]
	\centering
	\includegraphics[width=0.46\textwidth]{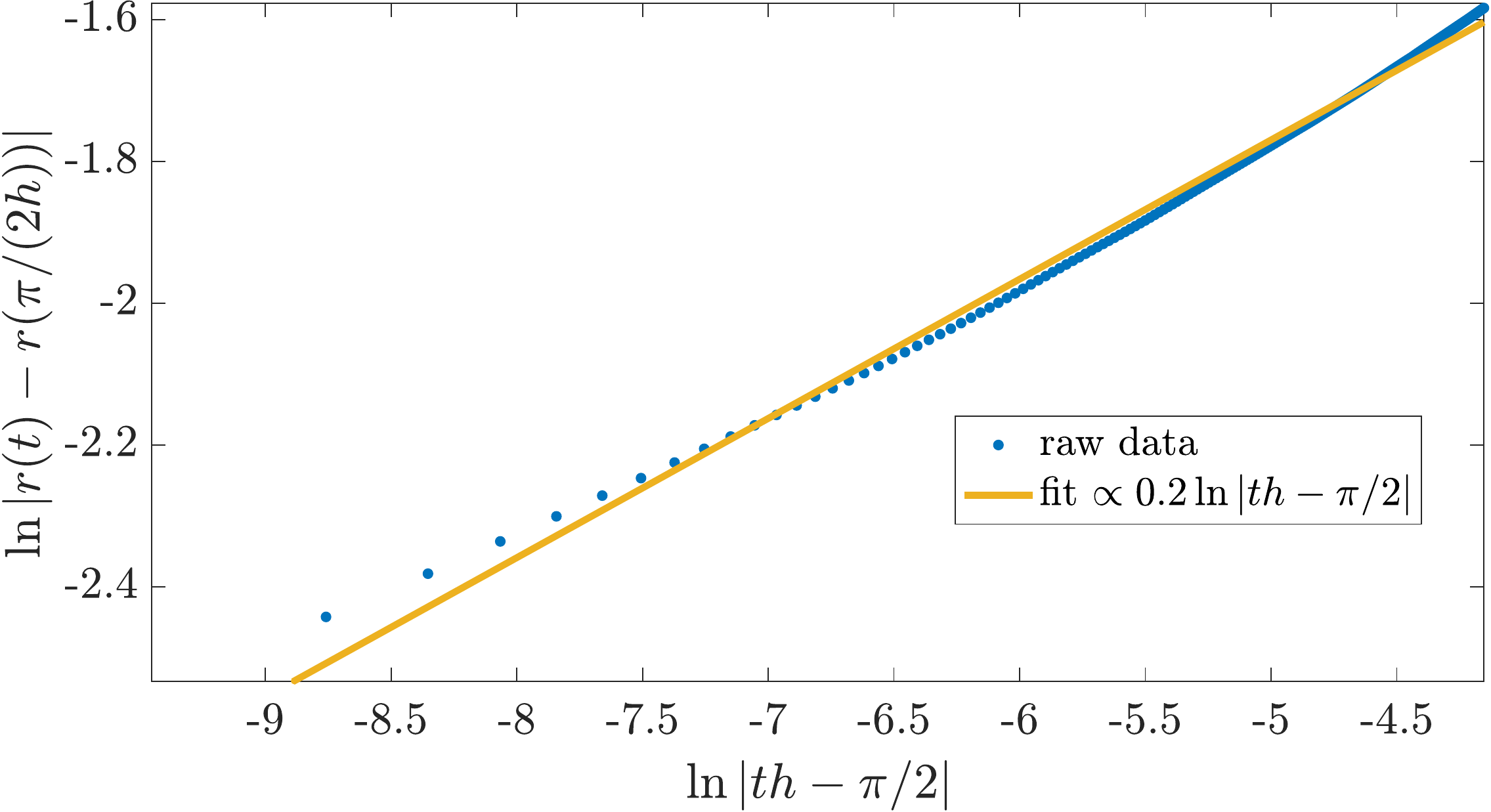}
	\caption{(Color online). Scaling of singularities of $r(t)$ in the vicinity of $t_n=\pi n/(2h)$ for model~\eqref{eq:randombond2} 
	indicating a power-law behavior $\propto|t-t_n|^\nu$ with $\nu \approx 0.2$.}
	\label{fit} 
\end{figure}

Ref.~\onlinecite{Gurarie2018} presented a detailed analysis of the singularities in~\eqref{eq:old} using analytic techniques.  That analysis is no longer available for the more intricate models of~\eqref{eq:randombond2} and~\eqref{eq:randombond3}, with random bonds instead of random fields.  Instead, we have studied singularities in $r(t)$ close 
to $t_n = \pi n/(2 h)$ in those models numerically. We find that these singularities are not in the universality class of~\eqref{eq:old}. Instead, they are of the power-law type
\begin{align} r(t) \sim \left| t - t_n \right|^\nu,\end{align}
with $\nu \approx 0.2$, as shown in Fig.~\ref{fit} for the model~\eqref{eq:randombond2}. The analysis of singularities in \eqref{eq:randombond3} 
also produces $\nu \approx 0.2$.

We have studied a number of other models of the type~\eqref{eq:classic} with one nonrandom interaction coupling and the rest random, and they all feature singularities supporting the conjecture stated above. It is possible, however, that in those other models the singularities belong to other
universality classes with distinct values of the exponent $\nu$. Studying this would be an interesting direction of further research.

Given that all spin-$1/2$ MBL systems map into~\eqref{eq:classic}, it is natural to expect that among them  systems can be found whose map into~\eqref{eq:classic} does not produce all random  and independent couplings. Those will necessarily feature singularities in their partition function ${\cal Z}(t)$. Identifying them however might not be easy. We propose~\eqref{eq:mb2} as the candidate model for this purpose. 
\begin{figure}[tb]
\centering
\centering
\hspace{-.25 cm}
\includegraphics[width=.46\textwidth]{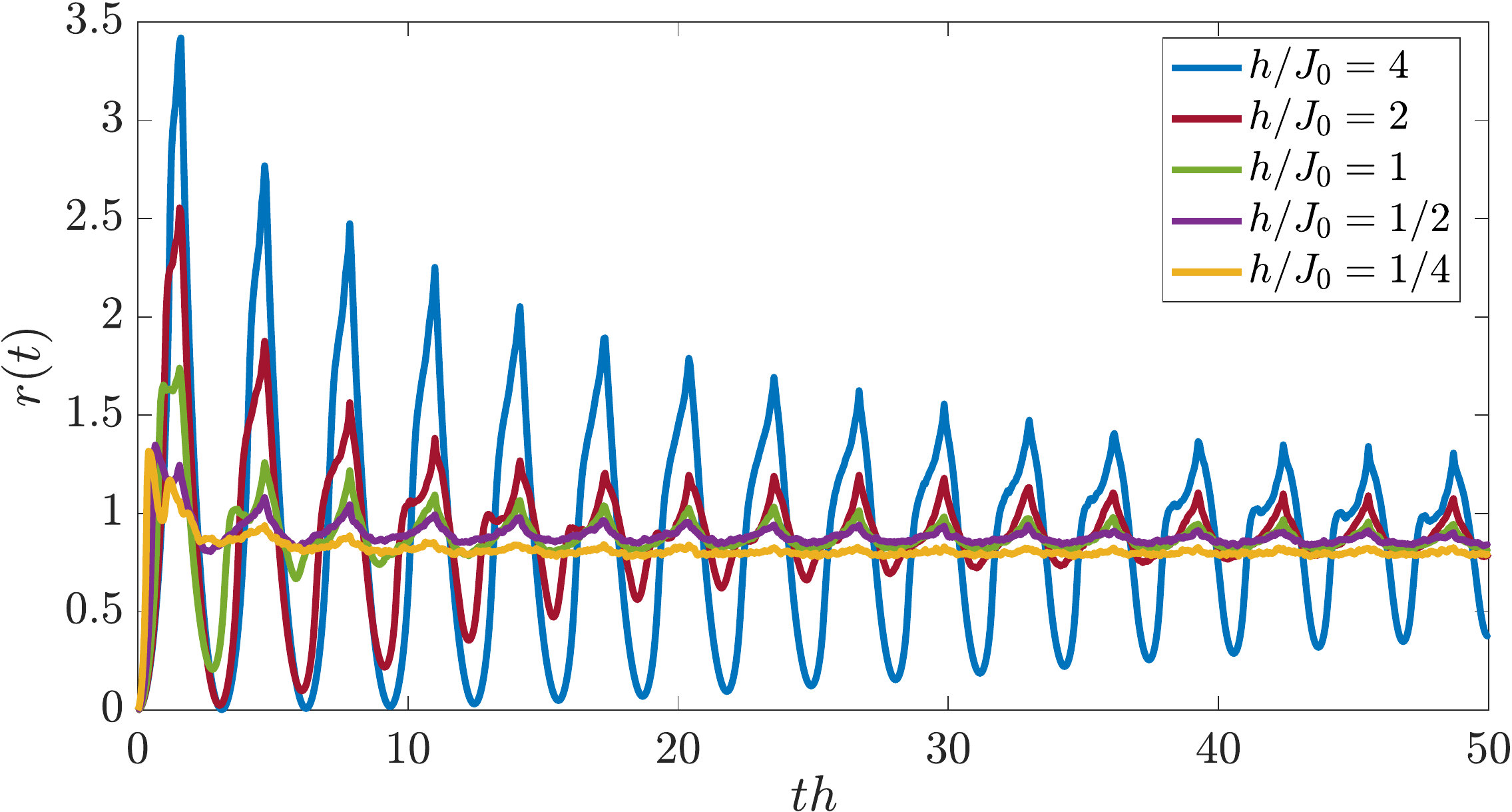}\\
\hspace{-.25 cm}
\includegraphics[width=.46\textwidth]{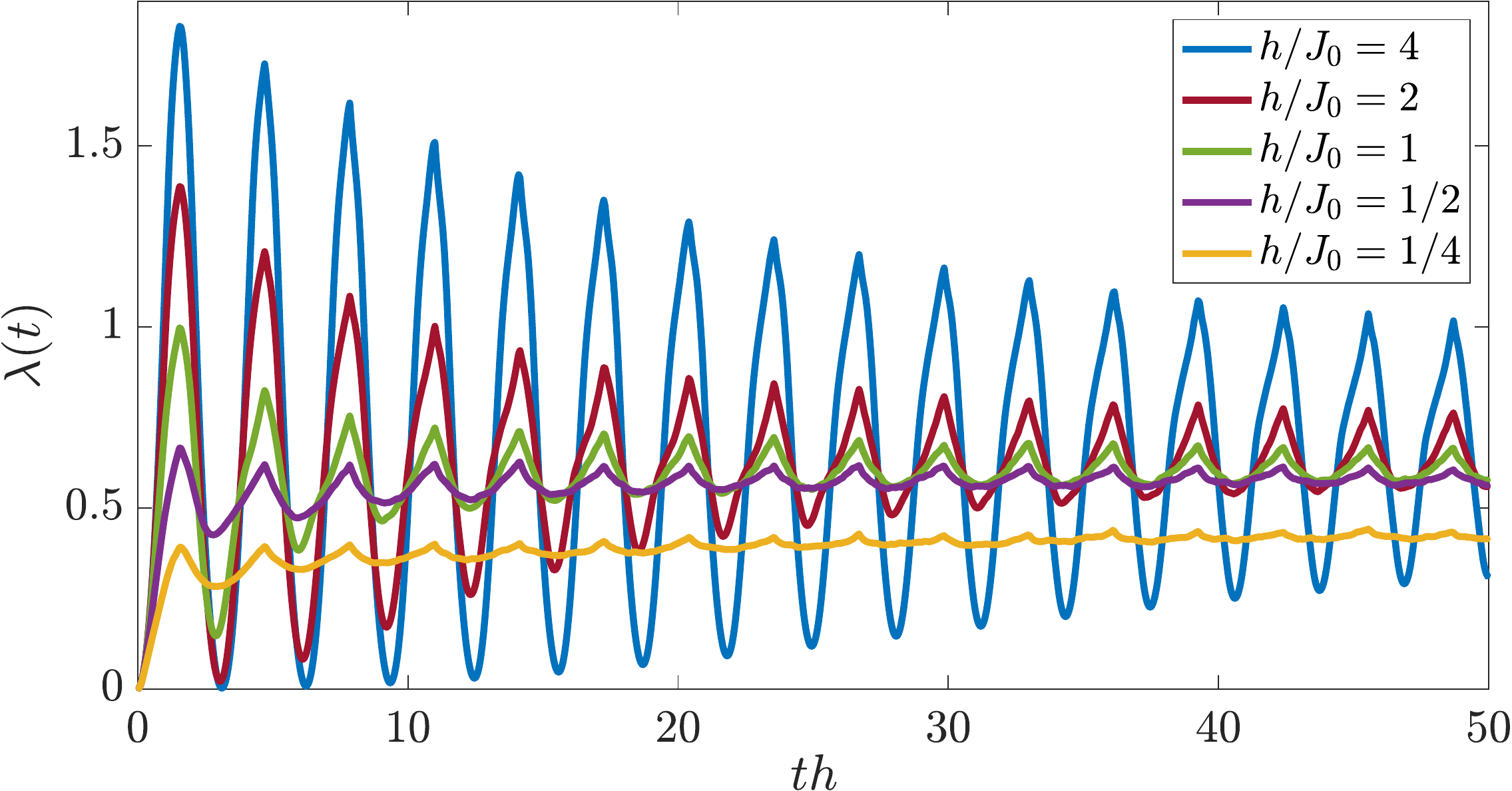}
\caption{(Color online). Partition-function return rate $r(t)$ given in~\eqref{eq:PFRR} (upper panel) and Loschmidt return rate $\lambda(t)$ given in~\eqref{eq:lerr} for the quantum model~\eqref{eq:mb2}, with each return rate averaged over $1000$ realizations of disorder. The initial state used for calculating $\lambda(t)$ is the fully $x$-polarized product state $\ket{X}$. It is clear that in the weak-disorder regime $h/J_0\gg1$ singularities appear in both return rates. The singularities seem to become less pronounced for $h/J_0\ll1$.}
\label{mbs} 
\end{figure}

\begin{figure}[tb]
	\centering
	\hspace{-.25 cm}
	\includegraphics[width=.46\textwidth]{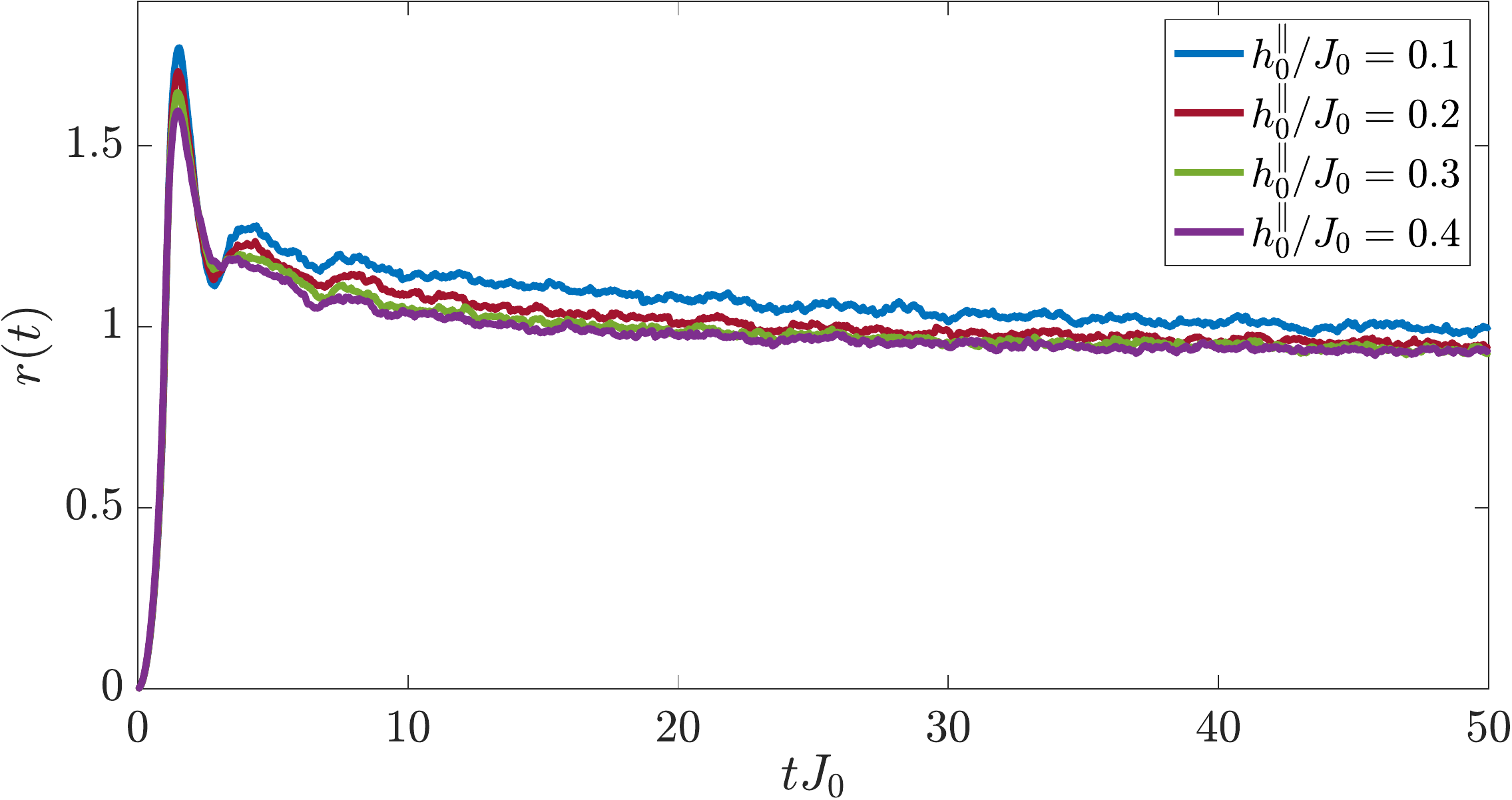}\\
	\hspace{-.25 cm}
	\includegraphics[width=.46\textwidth]{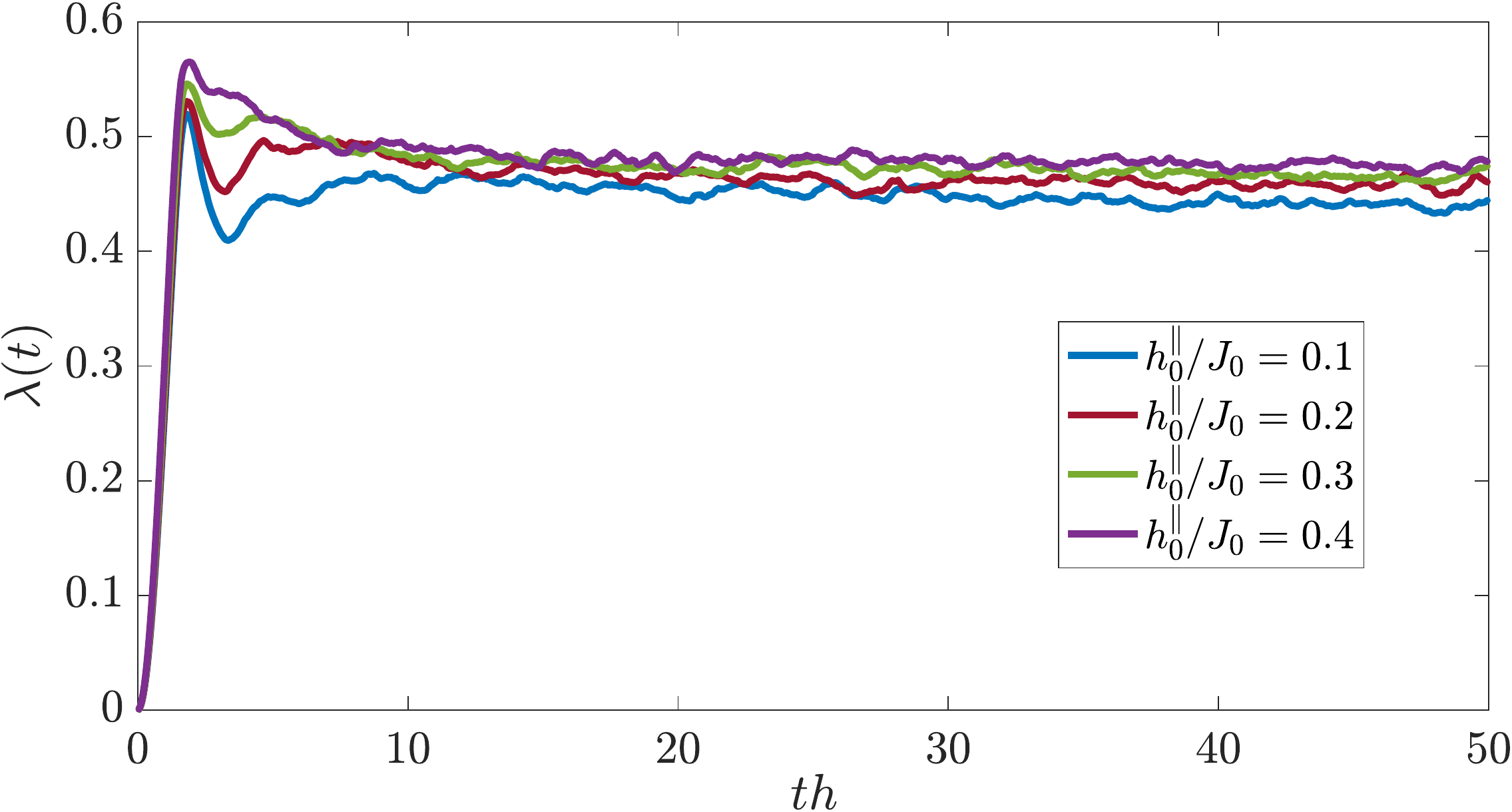}
	\caption{(Color online). Partition-function and Loschmidt return rates (top and bottom panels, respectively), each averaged over $1000$ disorder realizations, for the quantum model~\eqref{eq:mb1}, where $J_n,h_n^\perp\in[-J_0,J_0]$ and $h_n^\parallel\in[-h_0^\parallel,h_0^\parallel]$ are independent random variables. The initial state used for $\lambda(t)$ is $\ket{X}$. No singularities appear in either return rate for this model.}
	\label{nocusps} 
\end{figure}

Fig.~\ref{mbs} shows the result of evaluating the partition-function return rate $r(t)$~\eqref{eq:PFRR} and the Loschmidt return rate $\lambda(t)$~\eqref{eq:lerr}, calculated with $\ket{\psi} = \ket{X}$ being the state representing all spins polarized in the
$x$-direction, for the model~\eqref{eq:mb2} for $N=12$ sites by exact diagonalization, averaged over $1000$ realizations of disorder, necessarily due to the relatively short length of the chain. Random variables $J_n$ are taken to be uniformly distributed on the interval $[-J_0, J_0]$. Singularities are apparent in both return rates. Note that the periodicity of ${\cal Z}(t)$ is $\pi/h$. In fact, in the absence of disorder $J_0=0$, the return rate is easy to evaluate as
\begin{align} r(t) = -\frac{1}{N}\ln\frac{\left| 
	{\cal Z} \right|^2}{2^{2N}}= -  \ln \left|  \cos(h t) \right|.
\end{align}
We see that with disorder $r(t)$ retains the periodicity of its disorder-free version, but in addition clearly develops singularities, as desired. We emphasize that not all disordered many-body models exhibit singularities in the return rate as~\eqref{eq:mb2} does. For comparison, we
 can examine the return rates for the model~\eqref{eq:mb1}. Fig.~\ref{nocusps} clearly shows absence of any singularities in it  (Loschmidt return rate is calculated with the same $\ket{\psi} = \ket{X}$ as above). And indeed,
 we can argue that this model maps into an $l$-bit Hamiltonian with all random couplings.

We can further elucidate the model~\eqref{eq:mb2} by exploring the limiting cases of strong and weak disorder. In the strong disorder case, $h/J_0 \ll 1$, we can construct the $l$-bit Hamiltonian perturbatively; cf.~\eqref{eq:lbits}. Carrying out this procedure (see Appendix~\ref{sec:app}) in the second order
in $h/J_0$ does not produce any nonrandom couplings in~\eqref{eq:lbits}. This is in line with singularities disappearing for small $h/J_0$ in
Fig.~\ref{mbs}.

The limit of weak disorder, $h/J_0 \gg 1$, is much more subtle. At $J_0=0$ the model~\eqref{eq:mb2} results in many degenerate levels, separated by energy interval $2h$ (hence the periodicity of the partition function of $2\pi/(2h) = \pi/h$). Once disorder is turned on, each such level splits into a band. One can expect each band to be many-body localized. Indeed, the effective Hamiltonian within each band, which can be obtained
via second-order perturbation theory in $J_0/h$, is proportional to $J_0^2$, with its ratio to $J_0^2$ being $J_0$-independent. Thus the 
nature of the MBL eigenfunctions of this system does not depend on disorder strength as it is taken to zero, and no perturbation theory can be useful to analyze this phase in this limit.
The precise structure of the $l$-bit Hamiltonian is difficult to determine in this regime.  It is in this regime that the model~\eqref{eq:mb2} features singularities seen in Fig.~\ref{mbs}. We note the absence of the exact periodicity in Fig.~\ref{mbs}. That indicates that the appropriate $l$-bit Hamiltonian our model maps into cannot have simply a single nonrandom coefficient with the rest random and independent. That by itself
would produce a periodic-in-time $r(t)$ at large enough $t$. Rather the $l$-bit Hamiltoinan should have a more complicated structure involving correlations between its coefficients, going beyond the simple examples of~\eqref{eq:old}  or \eqref{eq:randombond1}.


Whether the singularities disappear at some critical value of $h/J_0$ or are present at all values albeit getting weaker as $h/J_0$ gets smaller 
cannot be explored with the methods currently available to us. The second scenario would imply that these singularities cannot be investigated
in perturbation theory over the small parameter $h/J_0$, consistent with the arguments in Appendix~\ref{sec:app}. Ultimately, MBL phases
are notoriously difficult to analyze using analytic techniques. It is therefore not surprising that we have to rely mostly on the numerics to analyze
our system.

\begin{figure}[tb]
	\centering
	\centering
	\hspace{-.25 cm}
	\includegraphics[width=.46\textwidth]{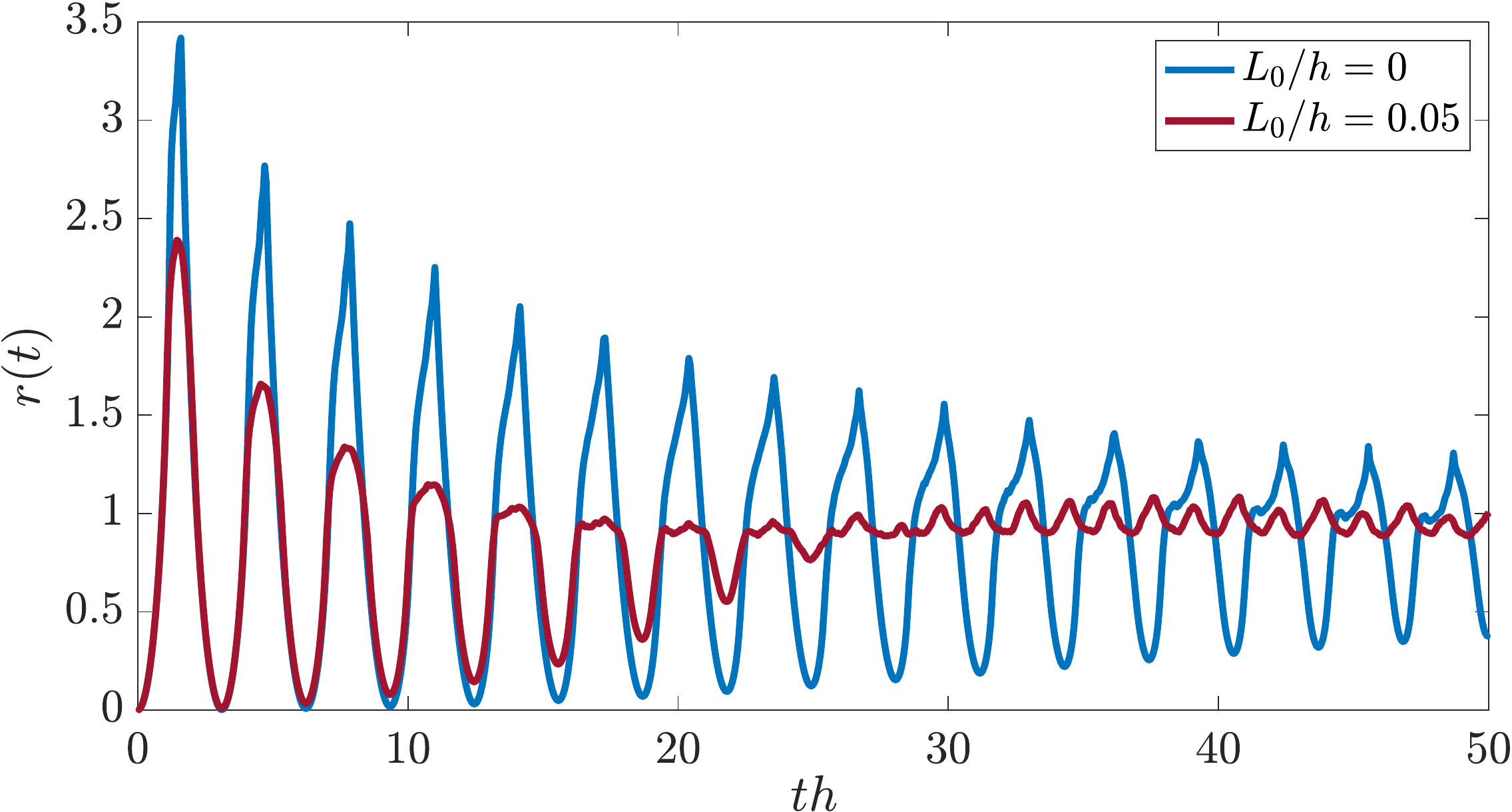}\\
	\hspace{-.25 cm}
	\includegraphics[width=.46\textwidth]{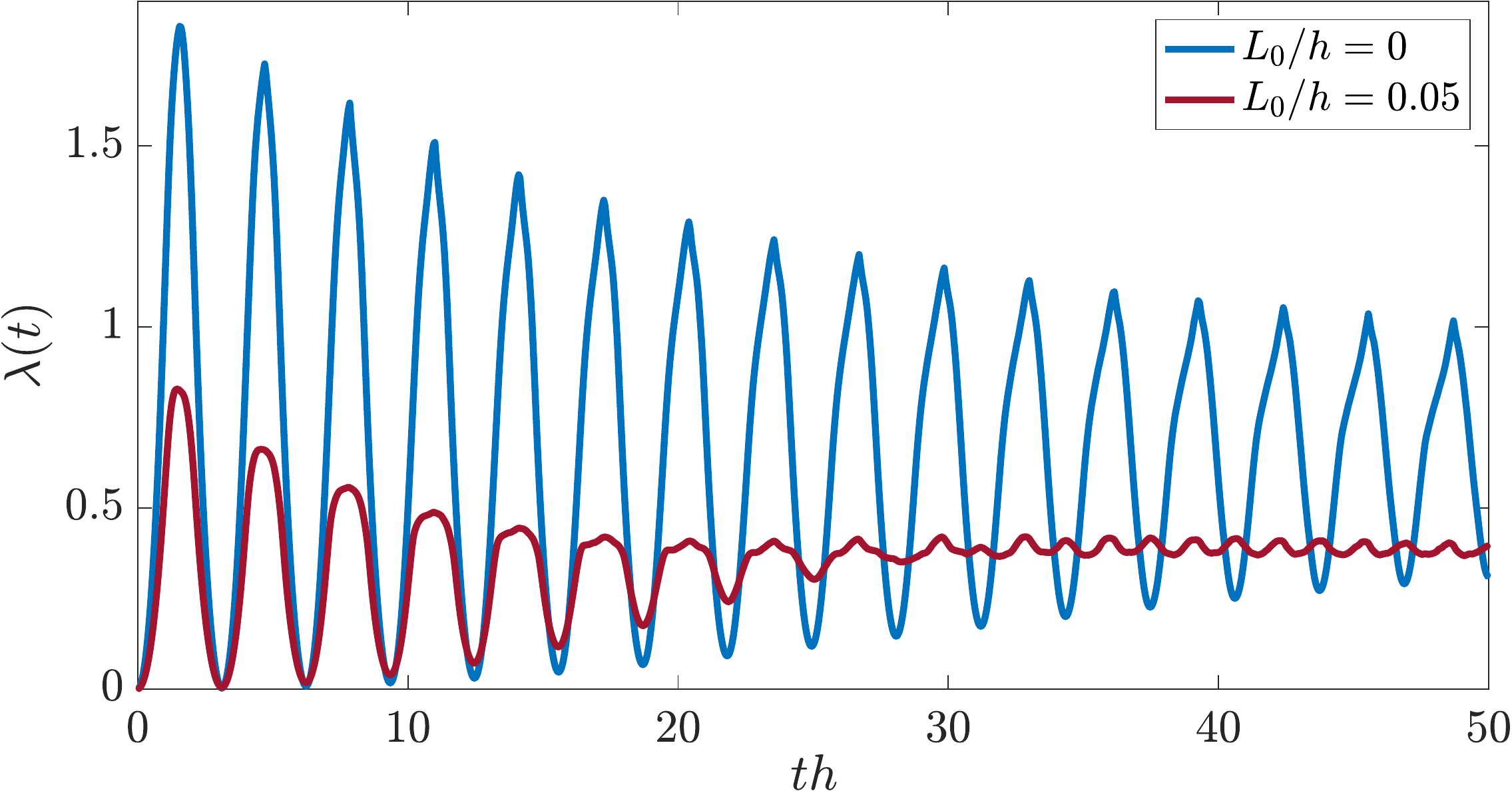}
	\caption{(Color online). Partition-function and Loschmidt return rates (top and bottom panels, respectively) for model~\eqref{eq:mb3} for fixed $h/J_0=4$ and two disorder strengths $L_0/h=0$ and $0.05$. A finite $L_0$ quickly washes out the singularities. Each return rate is averaged over $1000$ disorder configurations, with $\ket{X}$ the initial state used to compute $\lambda(t)$.}
	\label{fig:triple} 
\end{figure}
Finally, we observe that adding extra random terms to the Hamiltonian~\eqref{eq:mb2} can take us out of the class of MBL
Hamiltonians with singular return rates, even if the first term in~\eqref{eq:mb2} remains nonrandom. 
Consider for example
\begin{align}\label{eq:mb3} H = h \sum_{n=1}^N  \sigma^z_n + \sum_{n=1}^{N-2} \big(J_n \sigma^x_n \sigma^x_{n+1} \sigma^x_{n+2}+L_n \sigma^z_n \sigma^z_{n+1} \sigma^z_{n+2}\big),
\end{align} with $L_n$ random uniformly distributed on the interval $[-L_0, L_0]$. 
This Hamiltonian does not feature singularities in its return rate as shown in Fig.~\ref{fig:triple}, even for tiny $L_0/h = 0.05$. Clearly, only a certain subclass of MBL models represent systems with DQPT. In line with the arguments above, the Hamiltonian in~\eqref{eq:mb3} must map onto an
$l$-bit Hamiltonian with random and essentially uncorrelated interaction coefficients.

\section{Conclusion and outlook}\label{sec:conc}
In summary, we have investigated DQPT in quantum many-body systems with disorder. Using a mapping to $l$-bits, we have determined the conditions for which singularities appear in the return rate for such models, and presented candidates for which the return rate displays singularities for small to moderate disorder strength, and which may survive even at large disorder strength. Our results show that DQPT persist in quantum MBL systems, and are not restricted to clean quantum many-body models on which hitherto their investigation has been focused. In light of the search for an origin of DQPT, our conclusions confirm that a Landau equilibrium quantum phase transition is neither a necessary nor a sufficient condition for dynamical criticality to arise in the return rate. From an MBL point of view, this opens up several questions related to what MBL phases can imply about dynamical criticality. Even though this is well understood in traditional Landau phases and has been extensively studied in one-dimensional quantum Ising chains with various interaction ranges,\cite{Halimeh2017b,Halimeh2017a,Zunkovic2018,Zauner2017} two-dimensional models,\cite{Hashizume2018,Schmitt2015} and mean-field models,\cite{Homrighausen2017,Lang2018a,Lang2018b,Weidinger2017} little is known about how different MBL phases can alter the kind or presence of singularities in the return rate. In the other direction, one can wonder what the singularities in the return rate can tell us about the equilibrium, possibly MBL, phase. In Ref.~\onlinecite{Halimeh2018}, it is illustrated how one can determine the equilibrium physics, including Landau phases and the type of quasiparticle excitations in the spectrum of the quench Hamiltonian, directly from the return rate after a quantum-quench sweep, but no protocol was provided for discerning whether an equilibrium phase can be MBL. It would be interesting to further investigate this, and not necessarily just from the point of view of the return rate. Indeed, recently it has been shown how universal equilibrium scaling functions can be deduced at short times from spin-spin correlations after a quantum quench to the vicinity of a critical point.\cite{Karl2017} Additionally, it is worth mentioning that the MBL transition has been observed in experiments with interacting fermions in one-dimensional quasirandom optical lattices through the relaxation dynamics of the initial state.\cite{Schreiber2015} Our conclusions would in principle be amenable for observation in such experiments given that DQPT have also been observed in setups of spin-polarized fermionic atoms in driven optical lattices.\cite{Flaeschner2018}

Finally, questions remain on the possible universality classes of the DQPT observed in this work, and on general principles of 
mapping to $l$-bits Hamiltonians resulting in DQPT. We leave these open questions for future work.

\section*{Acknowledgments}
The authors acknowledge stimulating discussions with Dmitry Abanin, Maksym Serbyn, and Daniele Trapin. VG would also like to acknowledge support from the NSF Grant No. PHY-1521080.



\appendix
\section{$l$-bits and many-body localization}\label{sec:app}
We would like to further elucidate the structure of the partition function of the model
\be H = h \sum_{n=1}^N \sigma^z_n + \sum_{n=1}^{N-2} J_n \sigma^x_n \sigma^x_{n+1} \sigma^x_{n+2}
\ee
with $J_n$ independent random variables uniformly distributed on the interval $[-J_0, J_0]$. 
It is relatively easy to explore the regime where $\left| J_n \right| \gg h$ (if $J_0 \gg h$, the vast majority of $\left| J_n \right|$ will all be large). If $h=0$, then $x$ components of the spin operators are conserved,
\be 
\tau_j  = \sigma^x_j,
\ee and \be  [H, \tau_j]=0,\,\,\,\,\, \ \tau_j^2=1.
\ee
If $h>0$ but the ratios $h/\left| J_n \right| \ll 1$, then we can construct $\tau_j$ which commute with the Hamiltonian perturbatively. 
We write
\be \tau_j = \sigma^x_j + \tau^{(1)}_j + \tau^{(2)}_j + \dots,
\ee where each term above is suppressed compared to the previous one by an extra power of $h$. In the first order of perturbation theory 
we empose
\be h \sum_n [  \sigma^z_n, \sigma^x_j]  +  \sum_n J_n [ \sigma^x_n \sigma^x_{n+1} \sigma^x_{n+2}, \tau^{(1)}_j ]=0.
\ee
It is easy to solve this equation for $X^{(1)}_j$, to find
\begin{align} \tau^{(1)}_j =&\, \alpha_{j} \,  \sigma^z_j \sigma^x_{j+1} \sigma^x_{j+2} + \beta_{j} \, \sigma^x_{j-1} \sigma^z_j \sigma^x_{j+1} +
\cr & \gamma_{j} \,  \sigma^x_{j-2} \sigma^z_{j-1} \sigma^x_{j} + \delta_{j} \, \sigma^x_{j-2} \sigma^z_j \sigma^x_{j+2}. 
\end{align}
Here,
\begin{widetext}
\begin{align} \alpha_j &= \frac{J_j \left( J_j^2-J_{j-1}^2 - J_{j-2}^2 \right) h}{J_j^4+J_{j-1}^4+ J_{j-2}^4 - 2 J_j^2 J_{j-1}^2 - 2J_j^2 J_{j-2}^2 - 2 J_{j-1}^2 J_{j-2}^2}, \cr
\beta_j &= \frac{J_{j-1} \left( J_{j-1}^2-J_{j}^2 - J_{j-2}^2 \right)h }{J_j^4+J_{j-1}^4+ J_{j-2}^4 - 2 J_j^2 J_{j-1}^2 - 2J_j^2 J_{j-2}^2 - 2 J_{j-1}^2 J_{j-2}^2}, \cr
\gamma_j &= \frac{J_{j-2} \left( J_{j-2}^2-J_{j-1}^2 - J_{j}^2 \right) h}{J_j^4+J_{j-1}^4+ J_{j-2}^4 - 2 J_j^2 J_{j-1}^2 - 2J_j^2 J_{j-2}^2 - 2 J_{j-1}^2 J_{j-2}^2}, \cr
\delta_j &= \frac{2 J_j J_{j-1} J_{j-2} h }{J_j^4+J_{j-1}^4+ J_{j-2}^4 - 2 J_j^2 J_{j-1}^2 - 2J_j^2 J_{j-2}^2 - 2 J_{j-1}^2 J_{j-2}^2}. 
\end{align}
\end{widetext}
Note that in this order of perturbation theory,
\be \tau_j^2 \approx  \left( \sigma_j^x + \tau^{(1)}_j \right)^2  \approx  1,
\ee
up to terms of the second order in $h/J_n$. 

We would also like to construct further corrections to $\tau_j$ up to the second order of perturbation theory, satisfying
\be h \sum_n [  \sigma^z_n, \tau^{(1)}_j]  +  \sum_n J_n [ \sigma^x_n \sigma^x_{n+1} \sigma^x_{n+2}, \tau^{(2)}_j ]=0.
\ee
There are 28 terms in the expression for $\tau^{(2)}_j$. 24 of them anticommute with $\sigma^x_j$, and the remaining $4$ commute with it. 
The condition 
\be \tau_j^2 \approx \left( \sigma_j^x + \tau^{(1)}_j +\tau^{(2)}_j \right)^2  \approx  1,
\ee up to terms of the third order in $h/J_n$ implies that
\be \label{eq:four} \left( \tau^{(1)}_j \right)^2 +\sigma^x_j \tau^{(2)}_j + \tau^{(2)}_j  \sigma^x_j = 0.
\ee
The 24 terms in $\tau_j^{(2)}$ that anticommute with $\sigma^x_j$ do not contribute to this condition, but the remaining four terms do and can in fact be found from~\eqref{eq:four}. Those explicitly are
\begin{align} &-\left( \alpha_j \beta_j + \gamma_j \delta_j \right)\sigma^x_{j-1} \sigma^x_j \sigma^x_{j+2}, \cr 
&- \left( \alpha_j \delta_j + \beta_j \gamma_j \right) \sigma_{j-2}^x \sigma_j^x \sigma_{j+1}^x, \cr
&- \left( \alpha_j \gamma_j + \beta_j \delta_j \right) \sigma^x_{j-2} \sigma^x_{j-1} \sigma^x_j \sigma^x_{j+1} \sigma^x_{j+2},  \cr 
&- \oh\left( \alpha_j^2 + \beta_j^2 + \gamma^2_j + \delta_j^2 \right) \sigma^x_j. 
\label{eq:manyterms} \end{align}

Given the expression for the $l$-bits $\tau_j$ we can re-express the Hamiltonian in terms of them, according to~\eqref{eq:lbits}. The coefficients $K^{(s)}_{n_1 \dots n_s}$ can be found using the relation~\eqref{eq:lcoef}, where everything can now be written in terms of 
the original spin operators $\sigma$. 

Armed with the explicit expression for $\tau_j$ in terms of $\sigma$, we find that in this order of perturbation theory \be K^{(1)}_n=0. \ee
The coefficients $K^{(3)}_{n, n+1, n+2}$ which are equal to $J_n$ in the zeroth order of perturbation theory, acquire small corrections, which we won't reproduce explicitly here. 
At the same time, 
a new term 
\be K^{(3)}_{n, n+2, n+4}=\frac{h \delta_j}{2},
\ee
also appears. It can be evaluated entirely from the terms listed in \rfs{eq:manyterms}. Therefore, the partition function of the initial problem maps approximately into
\be {\cal Z}(t) = \sum_{\tau=\pm 1} e^{-i t \sum_n J_n \tau_n \tau_{n+1} \tau_{n+2} - i t \sum_n \delta_n \tau_n \tau_{n+2} \tau_{n+4}}.
\ee
There's no reason to expect that this would have any singularities as all coefficients here are random. And indeed, in the regime $h/J_0 \ll 1$ we do not see the singularities in Fig.~\ref{mbs}.


\bibliography{DCPTMBL}

\begin{thebibliography}{43}
\bibitem[{foo()}]{footnote}
\bibinfo{note}{One should note that in the context of MBL phases studied here,
  it is the return rate which should be averaged, not the spectral form factor,
  or alternatively a typical spectral form factor should be studied in place of
  the average one. However, typical spectral form factors are not well
  understood in the context of RMT\cite{Haake2015}}.
\end{thebibliography}


\end{document}